%Paper: hep-ph/9406258
%From: papavass@MAFALDA.PHYSICS.NYU.EDU (Joannis Papavassiliou)
%Date: Wed, 8 Jun 94 16:56:38 -0400

The paper is writen in Texsis, the macros are available in the SLAC Library;
There are 7 pages not included. Hard copies are available upon e-mail request
at papavass@mafalda.physics.nyu.edu
\magnification 1200
%\draft
\tenpoint
\superrefsfalse
\baselineskip = 0.333 truein \relax
\def\frac#1#2{{#1\over#2}}

\section {Introduction}

Issues of gauge invariance in the context of the Standard Model (SM)
have received considerable attention in recent years
\reference{D&SF}
G.~Degrassi and A.~Sirlin,
\journal Nucl. Phys.; B 383,73 (1992)
\endreference
{}.
To the extend that one is computing S-matrix elements such issues
are not important, since the propagators and vertices of the theory,
even though individually gauge dependent, conspire to give a gauge
invariant (g.i.) answer, order by order in perturbation theory
\reference{Abers}
E.~S.~Abers and B.~W.~Lee,
\journal Phys. Rep.;9C,1 (1973)
\endreference
{}.
On the other hand, when one attempts to extract
physical information out of individual parts of the S-matrix, the
gauge-invariance of the final answer is not guaranteed.
Such has been the case for example with
the conventionally defined electromagnetic form factors of particles
such as the W boson, the neutrino, and the top quark
\reference{Fujikawa}
K.~Fujikawa, B.~W.~Lee, and A.~I.~Sanda,
\journal Phys. Rev.; D 6,2923 (1972)
\endreference
\reference{Zepeda}
J.~L.~Lucio, A.~Rosado, and A.~Zepeda,
\journal Phys. Rev.; D 31,1091 (1985)
\endreference
\reference{*Grau}
A.~Grau and J.~A.~Grifols,
\journal Phys. Lett.;B116,233 (1986)
\endreference
\reference{*Sriv}
G.~Auriema, Y.~Srivastava, and A.~Widom,
\journal Phys. Lett.;B195,254 (1987)
\endreference
\reference{*DSM}
G.~Degrassi, A.~Sirlin, and W.~J~Marciano,
\journal Phys. Rev.; D 39,287 (1989)
\endreference
\reference{*Lahanas}
E.~N.~Argyres {\sl et al.},
\journal Nucl. Phys.; B391,23 (1993)
\endreference
, which when calculated in the
context of the $R_{\xi}$ gauges, turn out to be gauge dependent.
Similarly, the S, T, and U parameters
\reference{Peskin}
M.~Peskin and T.~Takeuchi,
\journal Phys. Rev. Lett; 65,964 (1990)
\endreference
,
which provide a convenient parametrization
of the oblique corrections of the SM, have also been shown to be
gauge dependent, and, except when calculated
in a restricted class of gauges,
even ultraviolet divergent
\reference{STU}
G.~Degrassi, B.~Kniehl, and A.~Sirlin,
\journal Phys. Rev.; D 48,R3963 (1993)
\endreference
{}.
In order to systematicaly address problems
of gauge-invariance, a method known as the
pinch technique (PT)
\reference{Kursunoglu}
J.~M.~Cornwall,
in Deeper Pathways in High Energy Physics, edited by B.~Kursunoglu,
A.~Perlmutter, and L.~Scott (Plenum, New York, 1977), p.683
\endreference
\reference{Bib}
J.~M.~Cornwall,
\journal Phys. Rev.; D 26,1453 (1982)
\endreference
, has been extensively employed.
In particular, a g.i
 definition of the S, T, and U parameters
was proposed \cite{STU},
and a general procedure for extracting g.i.
form factors has been developed in a series of papers
\reference{Klako}
J.~Papavassiliou,
\journal Phys. Rev.; D 41,3179 (1990)
\endreference
\reference {Kostas}
J.~Papavassiliou and K.~Philippides,
\journal Phys. Rev.; D 48,4255 (1993)
\endreference
\reference{Claudio}
J.~Papavassiliou and C.~Parrinello,
NYU/04/06, to appear in Phys.Rev.D
\endreference
{}.
More recently, an alternative formulation of the
renormalization procedure of the the SM in the framework of the PT
was proposed
\reference{Hagi}
K.~Hagiwara, S.~Matsumoto, and C.~S.~Kim,
{\sl Electroweak Radiative Corrections}, Plenary talk at the
14'th International Workshop on Weak Interactions and Neutrinos,
KEK preprint 93-108
\endreference
, and a renormalizable W-self energy was
constructed via the PT, in the context
of the {\sl unitary} gauge
\reference{P&S}
J.~Papavassiliou and A.~Sirlin,
{\sl Renormalizable W self-energy in the unitary gauge
via the pinch technique}, BNL-60241
\endreference
{}.
The PT is an algorithm that allows the construction
of modified g.i. n-point functions, through the
order by order rearrangement of
Feynman graphs contributing to a certain physical
and therefore ostensibly g.i. process,
such as an S-matrix element (Fig.1) or a Wilson loop.
In the simplest case of a four-fermion amplitude,
such as $q_{1}{\bar{q}}_{2}\rightarrow q_{1}{\bar{q}}_{2}$,
where $q_{1}$,$q_{2}$ are two
on-shell test quarks with masses $m_{1}$ and
$m_{2}$, respectively,
the application of the PT gives rise to a
g.i. gluon ($g$) self-energy ${\hat{\Pi}}_{\mu\nu}$,
and to g.i. $gq_{i}{\bar{q}}_{i}$
vertices ${\hat{\Gamma}}_{\mu}^{(i)}$
\reference{hats}
Throughout this paper we use hats to indicate
the g.i. quantities constructed via the PT
\endreference
{}.
The generalization of the PT from vector-like theories (such as QCD)
to the SM is technically and conceptually straightforward,
as long as one assumes that the external fermionic currents are conserved.
So, applying the PT
to a SM amplitude, such as $e^{-}\nu_{e}\rightarrow e^{-}\nu_{e}$,
with $m_{e}=m_{\nu}=0$
\reference{strictly}
Strictly speaking one only needs to assume that $m_{e}=m_{\nu}$
\endreference
, one finds a
$\xi$-independent self-energy for the $W$-boson, and a $\xi$-independent
$We^{-}\nu_{e}$ vertex. By analogy, the PT applied on a neutral process
such as $e^{+}e^{-}\rightarrow e^{+}e^{-}$,
with $m_{e}=0$, gives rise to a
$\xi$-independent $Z$ self-energy, and $Ze^{+}e^{-}$ vertex
\reference{D&S}
G.~Degrassi and A.~Sirlin,
\journal Phys. Rev.; D 46,3104 (1992)
\endreference
{}.
In addition to being g.i.,
the vertices ${\hat{\Gamma}}_{\mu}$
constructed in all of the aforementioned cases satisfy
simple QED like Ward identities of the form
$q^{\mu}{\hat{\Gamma}}_{\mu}=S^{-1}_{i}(p)-S^{-1}_{j}(p+q)$,
where $S^{-1}_{i}$
is the inverse propagator of the $i$-th fermion, calculated in the
Feynman gauge ($\xi=1$)
\reference{remark}
Clearly, in the QCD case, as well as in the neutral processes,
we have that $i=j$
\endreference
{}.

The situation becomes more involved if one decides to
consider non-conserved external fermionic currents.
In the neutral current case this is equivalent to assuming that
$m_{i}\not= 0$, whereas in the charged current case $m_{i}\not= m_{j}$.
 The reason for the complications is that,
when evaluating such an amplitude, unlike the
conserved current cases,
 one has to also take into account contributions from
longitudinal degrees of freedom,
which originate either from the
gauge boson propagator, or from the unphysical Goldstone bosons,
whose Yukawa couplings are
proportional to the mass difference of the external fermions.
So, in addition to the usual $W^{+}W^{-}$
self-energy $\Pi_{\mu\nu}$ one has to consider the
$\phi^{+}\phi^{-}$ self-energy $\Omega$ of the Goldstone bosons, and the
$W^{+}\phi^{-}$ and $\phi^{+}W^{-}$ mixing terms
$\Theta_{\mu}$ and $\Theta_{\nu}$, respectively
Similarly, in addition to the $W\bar{f_{1}}f_{2}$ vertex $\Gamma_{\mu}$,
the $\phi\bar{f_{1}}f_{2}$ vertex $\Lambda$ must be considered.
Just like $\Pi_{\mu\nu}$ and $\Gamma_{\mu}$,
these new quantities are gauge dependent at one-loop.

In this paper we concentrate on four-fermion amplitudes, with
non-conserved external charged currents. In particular, we consider the
process $e^{-}\nu_{e}\rightarrow e^{-}\nu_{e}$,
with $m_{\nu}=0$, but with $m_{e}\not= 0$, and show that:

(a)~~ Proper use of the PT gives rise at one-loop order to
g.i. self-energies ${\hat{\Pi}}_{\mu\nu}$,
${\hat{\Theta}}_{\mu}$, ${\hat{\Theta}}_{\nu}$, and $\hat{\Omega}$,
and g.i. vertices ${\hat{\Gamma}}_{\mu}$ and $\hat{\Lambda}$
(Fig.2).
This task is technically rather involved, mainly because of two reasons.
First, the propagator-like pinch contributions (isolated from vertex
and box diagrams) are not just added to $\Pi_{\mu\nu}$
(as in the case of the conserved currents),
 but must be judiciously  allotted among the
$\Pi_{\mu\nu}$, $\Theta_{\mu}$, $\Theta_{\nu}$ and $\Omega$,
in order to construct the g.i.
${\hat{\Pi}}_{\mu\nu}$, ${\hat{\Theta}}_{\mu}$, ${\hat{\Theta}}_{\nu}$
and $\hat{\Omega}$.
Second, new propagator-like pinch contributions, not encountered previously,
appear now, originating from vertex graphs with an incoming
$\phi^{\pm}$, shown in Fig.7(d,e,f,g,h).
All such contributions must be identified and isolated from the respective
vertex graphs, since their contribution is essential for the success of the
program.

(b)~~ We show that the g.i. quantities ${\hat{\Pi}}_{\mu\nu}$,
${\hat{\Theta}}_{\mu}$, ${\hat{\Theta}}_{\nu}$, and $\hat{\Omega}$,
obtained via the PT, are
related by the following Ward identities:
$$
q^{\mu}q^{\nu}{\hat{\Pi}}_{\mu\nu}
-2Mq^{\mu}{\hat{\Theta}}_{\mu}+M^{2}\hat{\Omega}=0 ~~,
\EQN WIa$$
$$
q^{\mu}{\hat{\Pi}}^{\mu\nu}-M{\hat{\Theta}}_{\nu}=0 ~~,
\EQN WIb$$
$$
q^{\mu}{\hat{\Gamma}}_{\mu}-M\hat{\Lambda}=0 ~~.
\EQN WIc$$
\Eq{WIa}
is the generalization of the Passarino-Veltman
Ward identities
\reference{P&V}
G.~Passarino and M.~Veltman,
\journal Nucl. Phys.; B 160,151 (1979)
\endreference
to {\sl individually} g.i self-energies.
On the other hand, \Eq{WIb} and \Eq{WIc} have no analog in the context
of conventionally defined self-energies and vertices, and their validity is
particular to the PT.
The derivation of the above Ward identities is based solely on
the requirement that the S-matrix is
g.i. to a given order in perturbation theory, and knowledge of the
explicit closed form of the quantities involved is {\sl not} necessary.

c)~~ Use of the aforementioned Ward identities
enables the decomposition of the S-matrix into
{\sl individually} g.i. transverse and longitudinal
propagator-like, vertex-like, and box-like contributions.
So, to one loop order the S-matrix may be cast in the form
$$
S= \sum_{i=1}^{3}[{\hat{T}}^{tr}_{i}+{\hat{T}}^{lon}_{i}]~~,
\EQN TrLon$$
where $\hat{T_{1}}(t)$ is the propagator-like part,
$\hat{T_{1}}(t,m_{e})$ is the vertex-like part, and
$\hat{T_{3}}(t,s,m_{e})$ the box-like part,
and $t$ and $s$ are the usual Mandelstam variables.
In addition, the mixing terms between $W$ and $\phi$ disappears after this
rearrangement, thus leading to a one-loop generalization of the
well known tree-level property of the $R_{\xi}$ gauges \cite{Fujikawa}.

There are several reasons,
both theoretical and phenomenological,
motivating the application of the PT to the
case of non-conserved currents.
To begin with, this is obviously the physically relevant limit, since
quarks and leptons are endowed with different masses.
{}From the phenomenological point of view, treating the external fermions as
massless is an acceptable approximation, as long as their masses are
considerably lighter than the characteristic mass scale of the W and Z bosons.
Such an assumption clearly breaks down
when the external fermions involve the top quark.
The computation of top quark form-factors, for example,
for the process $e^{+}e^{-}\rightarrow t\bar{t}$
\reference{Sona}
D.~Atwood and A.~Soni,
\journal Phys. Rev.; D 45,2405 (1992)
\endreference
requires the use of the
PT in order to guarantee the gauge-invariance of the
final answer \cite{Claudio}.
Another interesting case is the
study of CP violation in semileptonic top decays of the form
$t\rightarrow b\tau^{+}\nu_{\tau}$
\reference{Sonb}
D.~Atwood, G.~Eilam, A.~Soni, R.~P.~Mendel, and R.~Migneron,
\journal Phys.Rev.Lett.;70,1364 (1993)
\endreference
{}.
In the context of the Weinberg model for CP violation
\reference{Weinberg}
S.~Weinberg,
\journal Phys.Rev.Lett.;37,657 (1976)
\endreference
the physical observables considered
 are proportional to the interference term
between graphs involving a $W$ boson and graphs with an extra
heavy charged Higgs. Due to helicity mismatches, only the
longitudinal parts of the usual SM graphs  contribute to these
observables, so that the entire effect is proportional to
the product $m_{t}m_{\tau}$ of the masses of the two heavy
external fermions.
The decomposition described in (c)
not only organizes such
calculations in a manifestly g.i. way, but it also eliminates
a great deal of labor
%%%% Here I removed Sirlin
\reference{CPa}
J.~Papavassiliou,
{\sl Issues of CP violation in top quark decays in
the Weinberg Model}, in preparation
\endreference
, since it isolates and discards a g.i. transverse
subset of Feynman graphs
\reference{CPb}
In the original analysis of \cite{Sonb} the only one-loop contributions
considered were due
the imaginary parts of fermionic loops in the W self-energy.
Of course, such contributions are automatically g.i.
However, as explained in \cite{CPa},
the observables studied in \cite{Sonb} also
receive contributions from the imaginary parts of graphs
containing gauge bosons, not considered in \cite{Sonb}. When dealing
with such contributions, issues of gauge invariance become
important, and the formalism developed in the
present paper turns out to be particularly useful
\endreference
{}.
In addition, complicated two-loop calculations can be facilitated
from the use of the PT. In such cases, regardless of any assumptions about
the external fermion masses, off-shell $WW$, $\phi\phi$ and
mixed self-energies
appear as subsets of the full calculation, and knowledge of
Ward identities between them can simplify the analysis, and provide
useful checks for the algorithms employed
\reference{WSB}
G.~Weiglein, R.~Scharf, and M.~B\"ohm,
\journal Nucl. Phys.; B 416,606 (1994)
\endreference
{}.
In a different
theoretical framework, the results described above
are a further step toward the construction of g.i. Schwinger-Dyson (S.D.)
equations for non-Abelian gauge theories,
either with elementary Higgs particles or with dynamical symmetry breaking.
In fact, the derivation of S.D. equations which
would be g.i. order by order in a dressed loop expansion has been the
original motivation for the introduction of the PT
\cite{Kursunoglu},\cite{Bib}. Such a treatment involves off-shell
self-energies and vertices; their extraction
from four-fermion amplitudes, such as the ones described above,
is an expeditious way for enforcing their
gauge independence.
The details of this general
approach have been already discussed in a series of
papers
\reference{prog}
J.~M.~Cornwall,
\journal Phys. Rev.; D 38,656 (1988)
\endreference
\reference{*C&P}
J.~M.~Cornwall and J.~Papavassiliou,
\journal Phys. Rev.; D 40,3474 (1989)
\endreference
\reference{*4g}
J.~Papavassiliou,
\journal Phys. Rev.; D 47,4728 (1993)
\endreference
{}.
Even though the program is not yet completed, the present analysis
generalizes the results of the simple $SU(2)$ toy model presented in
\cite{Klako} to the more
complicated and realistic case of the SM. In this context,
the decomposition of the S-matrix mentioned at (c) assures
the renormalizability of the resulting S.D. equations.

It is important to emphasize that the closed form of the g.i.
self-energies
obtained by the application of the S-matrix pinch technique does {\sl not}
depend on the particular process employed,
as one can easily verify by explicit calculations. On the other
hand, the g.i.
vertices ${\hat{\Gamma}}_{\mu}$ and $\hat{\Lambda}$ clearly
depend on the specific process, through the fermion masses
entering into their loops.
The Ward identity of \Eq{WIc}, which these
vertices satisfy, are however {\sl process-independent}.
So, for example, the
$We^{-}\bar{\nu}$ and $\phi e^{-}\bar{\nu}$ vertices are different
from the $Wt\bar{b}$ and $\phi t\bar{b}$
vertices, but both sets satisfy \Eq{WIc}.
So, in that sense, the Ward identities presented here
reflect an intrinsic process-independent property of the theory.

The paper is organized as follows: In section 1 we briefly review
some of the features of the PT, which are relevant to our purposes.
In section 2 we present a detailed analysis of the modifications necessary
for the application of the PT in the context of the SM with
non-conserved external currents.
In section 3 an explicit example is presented, where
the formalism developed in the previous section is applied.
In section 4 we derive the Ward identities
 for the g.i self-energies and vertices,
and rearrange the S-matrix into individually g.i. transverse
and longitudinal structures. In section 5 we explicitly prove the
Ward identities derived
in the previous section, to one-loop order. Finally, in section 5 we present
our conclusions.

\section { The Pinch Technique}
The simplest example that demonstrates how the PT works is the gluon
two point function (propagator).
Consider the $S$-matrix
element $T$ for the elastic scattering of two fermions of masses
$m_{1}$ and $m_{2}$. To any order in perturbation theory $T$ is independent
of the gauge fixing parameter $\xi$, defined by the free
gluon propagator
$$
\Delta_{\mu\nu}(q) = \frac{-i}{q^2}
[g_{\mu\nu}-(1-\xi)\frac{q_{\mu}q_{\nu}}{q^2}] .
\EQN GaugeProp$$
On the other hand, as an explicit calculation shows,
the conventionally defined proper self-energy (collectively
depicted in graph 1a)
depends on $\xi$. At the one loop level this dependence is canceled by
contributions from other graphs, like 1b and 1c, which,
at first glance, do not seem to be
propagator-like.
That this cancellation must occur and can be employed to define a
g.i. self-energy, is evident from the decomposition:
$$
T(s,t,m_{1},m_{2})= T_{1}(t) + T_{2}(t,m_{1},m_{2})+T_{3}(s,t,m_{1},m_{2})
\EQN S-matrix$$
where the function $T_{1}(t)$ depends only on the Mandelstam variable
$t=-({\hat{p}}_{1}-p_{1})^{2}=-q^2$,
 and not on $s=(p_{1}+p_{2})^{2}$ or on the
external masses.
Typically, self-energy, vertex, and box diagrams
contribute to $T_{1}$, $T_{2}$, and $T_{3}$, respectively.
Moreover, such contributions are $\xi$ dependent. However, as the sum
$T(s,t,m_{1},m_{2})$ is g.i., it is easy to show that
\Eq{S-matrix} can be recast in the form
$$
T(s,t,m_{1},m_{2})=
{\hat{T}}_{1}(t) + {\hat{T}}_{2}(t,m_{1},m_{2})+
{\hat{T}}_{3}(s,t,m_{1},m_{2}) ,
\EQN S2-matrix$$
where the ${\hat{T}}_{i}$ ($i=1,2,3$) are {\sl separately} $\xi$-independent.
  The propagator-like parts of graphs like 1e and 1f,
which enforce the gauge independence of $T_{1}(t)$,
 are called  "pinch parts".
The pinch parts emerge every time a gluon propagator or an elementary
three-gluon vertex contribute a longitudinal $k_{\mu}$ to the original
graph's numerator. The action of such a term is
to trigger an elementary
Ward identity of the form
$$\eqalign{
k^{\mu}\gamma_{\mu} \equiv & \slashchar{k} = (\slashchar{p}+
\slashchar{k}-m)-(\slashchar{p}-m)\cr
=& S^{-1}(p+k)-S^{-1}(p)\cr}
\EQN BasicPinch$$
once it gets contracted with a $\gamma$ matrix.
The first term on the right-hand side of \Eq{BasicPinch} will remove the
internal fermion propagator - that is a "pinch"
- whereas $S^{-1}(p)$ vanish
on shell.
Returning to the decomposition of \Eq{S-matrix}, the function
${\hat{T}}_{1}$ is
g.i. and may be identified with the contribution
of the new propagator.
We can construct the new propagator, or equivalently
${\hat{T}}_{1}$, directly from
the Feynman rules. In doing so it is evident that any value for the gauge
parameter $\xi$ may be chosen, since
${\hat{T}}_{1}$, ${\hat{T}}_{2}$, and ${\hat{T}}_{3}$ are
all independent of $\xi$. The simplest of all covariant gauges is
certainly the Feynman gauge ($\xi = 1$),
 which removes the
longitudinal part of the gluon propagator. Therefore, the only possibility
for pinching in four-fermion amplitudes arises from the
four-momentum of the three-gluon vertices, and the only
 propagator-like contributions come from graph 1b.

To explicitly calculate the pinching contribution of a graph such as 1b
it is convenient to decompose the vertex in the following way
(omitting a factor $i\epsilon_{abc}$)
$$
\Gamma_{\mu\nu\alpha}=\Gamma^{P}_{\mu\nu\alpha} + \Gamma^{F}_{\mu\nu\alpha}
\EQN tHooft$$
with
$$\eqalign{
&\Gamma^{P}_{\mu\nu\alpha} = (q+k)_{\nu}g_{\mu\alpha}
 + k_{\mu}g_{\nu\alpha}\cr
&\Gamma^{F}_{\mu\nu\alpha} = 2q_{\mu}g_{\nu\alpha} -
 2q_{\nu}g_{\nu\alpha} - (2k+q)_{\alpha}g_{\mu\nu}\cr}
\EQN GammaF$$

Now $\Gamma^{F}_{\mu\nu\alpha}$ satisfies a Feynman-gauge Ward identity:
$$
q^{\alpha}\Gamma^{F}_{\mu\nu\alpha} = [k^2-(k+q)^2]g_{\mu\nu}
\EQN FeynWard$$
where the RHS is the difference of two inverse propagators in the
Feynman gauge.
As for $\Gamma^{P}_{\mu\nu\alpha}$ (P for "pinch"), it gives rise to pinch
parts when contracted with $\gamma$ matrices
$$\eqalign{
&g_{\mu\alpha}(\slashchar{q}+\slashchar{k})=
 ig_{\mu\alpha}[S^{-1}(p+q)-S^{-1}(p-k)]\cr
&g_{\nu\alpha}\slashchar{k}=
ig_{\nu\alpha}[S^{-1}(p)-S^{-1}(p-k)]\cr}
\EQN Pinch2$$
Now both
$S^{-1}(p+q)$ and $S^{-1}(p)$ vanish on shell, whereas the two terms
proportional to $S^{-1}(p-k)$ pinch out the internal fermion propagator
in graph 1b, and so we are left with two "pinch" (propagator-like)
parts and one "regular" (purely vertex-like) part, namely
$$
pinch~part=2\times(-\frac{1}{2}c_{A})ig^2
\int_{n}\frac{1}{k^2(k+q)^2}(g^{\mu\rho}\gamma_{\rho})
\EQN HalfPinch$$
$$
regular~part=(\frac{1}{2}c_{A})ig^2
\int_{n}\frac{\gamma^{\rho}S(p-k)
\gamma^{\sigma}{\Gamma}^{F}_{\rho\sigma\mu}}
{k^2(k+q)^2}
\EQN RegularVertex$$
with $c_{A}$ the Casimir operator for the adjoint representation
[ $c_{A}=N$ in $SU(N)$ ]
and a factor of 2 for
the two pinching terms of \Eq{Pinch2},
and
$\int_{n}\equiv {\mu}^{4-n}\int \frac{d^{n}k}{(2\pi)^{n}}$.
When we add to the usual propagator graphs (in the Feynman gauge)
 the pinch contributions of \Eq{HalfPinch}, together
with an equal contribution coming from the mirror-image graph of 1b,
we find the new g.i. self-energy
${\hat{\Pi}}_{\mu\nu}(q)$, given by
$$
{\hat{\Pi}}_{\mu\nu}(q)= (q^{2}g_{\mu\nu} - q_{\mu}q_{\nu})\hat{\Pi}(q)
\EQN GaugInvProp$$
where
$$
\hat{\Pi}(q)= - bg^2\ln(\frac{-q^2}{\mu^2})
\EQN RunnCoupl$$
and $b= \frac{11N}{48\pi^2}$
the coefficient in front of $- g^3$ in the usual one loop $\beta$ function.
After all pinch contributions from graph 1b have been allotted to the new
gluon self-energy, the rest e.g.
 the expression in \Eq{RegularVertex}, is genuinely vertex-like and must be
added to the usual QED-like graphs. The final expression for the g.i.
vertex ${\hat{\Gamma}}^{a}_{\alpha}$ is given by
$$
{\hat{\Gamma}}^{a}_{\mu}=i\tau_{\alpha}g^2
\Biggl\lbrack(\frac{c_{A}}{2})
\int_{n}
\frac{\gamma^{\rho}S(p-k)
\gamma^{\sigma}{\Gamma}^{F}_{\rho\sigma\mu}}
{k^2(k+q)^2}
+(\frac{c_{A}}{2}-\frac{Nd_{f}}{c_{f}})
\int_{n}\frac{\gamma^{\rho}S(p+q-k)\gamma_{\mu}
S(p-k)\gamma_{\rho}}{k^2}\Biggr\rbrack
\EQN GaugeInvVert$$
where $\tau_{\alpha}$ is the fermion representation matrix, $d_{f}$ its
dimension, and $c_{f}$ its Dynkin index.
Acting with $q^{\mu}$
on ${\hat{\Gamma}}^{a}_{\mu}$ of \Eq{GaugeInvVert}, and using
\Eq{FeynWard}, we obtain the following QED-like
Ward identity:
$$
q^{\mu}{\hat{\Gamma}}^{a}_{\mu}=
\tau^{a}[\Sigma(p)-\Sigma(\hat{p})]
\EQN WardIden$$
with $\Sigma$ the quark self-energy in the Feynman gauge, and
$\hat{p}=p+q$.

\section {PT in the $R_{\xi}$ gauges with non-conserved currents.}

In this section we discuss the technical subtleties encountered
in the application of the PT
when the external fermions have different masses.
 The main motivation is to construct via the PT
g.i one-loop $WW$ $\phi W$, $W\phi$, and $\phi\phi$ self energies,
which we will call ${\hat{\Pi}}_{\mu\nu}$,
 ${\hat{\Theta}}_{\mu}$,
${\hat{\Theta}}_{\nu}$, and $\hat{\Omega}$ respectively, as well as
g.i. $W\bar{f_{1}}f_{2}$ and $\phi\bar{f_{1}}f_{2}$
vertices, which we call ${\hat{\Gamma}}_{\mu}$ and $\hat{\Lambda}$,
respectively (Fig.2).

Before we  proceed, we record some useful formulas.
The tree-level vector-meson propagator $\Delta_{\mu\nu}^{i}(q)$
 in the $R_{\xi}$ gauges is given by \cite{Fujikawa}
$$
\Delta_{\mu\nu}^{i}(q)= [g_{\mu\nu} - (1-\xi_{i})\frac{q_{\mu}q_{\nu}}
{q^{2}-\xi_{i}M^{2}_{i}}]\frac{-i}{q^{2}-M^{2}_{i}} ~~,
\EQN Wprop$$
with $i=w,z,\gamma$, and $M_{\gamma}=0$.
Its inverse
 ${[\Delta_{i}^{-1}(q)]}^{\mu\nu}$
is given by
$$
{[\Delta_{i}^{-1}(q)]}^{\mu\nu}= i[(q^{2}-M^{2}_{i})g^{\mu\nu}
- q^{\mu}q^{\nu}+ \frac{1}{\xi_{i}}q^{\mu}q^{\nu}]~~.
\EQN InvKsi$$
The propagators $D^{i}(q)$ of the
unphysical Goldstone bosons are
given by
$$
D^{i}(q)=\frac{i}{q^{2}-\xi_{i} M^{2}_{i}} ~~,
\EQN Gold$$
and explicitly depend on $\xi_{i}$. On the other hand,
the propagators of the fermions (quarks and leptons), as well as the
propagator
of the physical Higgs particle
are $\xi_{i}$-independent at tree-level.
The following identities, which hold for {\sl every} value of
the gauge fixing parameters $\xi_{i}$,
will frequently be used:
$$
\Delta_{\mu\nu}^{i}(q)=
{\cal D}_{\mu\nu}^{i}(q)-\frac{q_{\mu}q_{\nu}}{M^{2}_{i}}D^{i}(q)~~,
\EQN Id1$$
where
$$
{\cal D}_{\mu\nu}^{i}(q)= [g_{\mu\nu} - \frac{q_{\mu}q_{\nu}}
{M^{2}_{i}}]\frac{-i}{q^{2}-M^{2}_{i}}
\EQN UnitaryProp$$
is the the $W$ and $Z$ propagator in the unitary gauge
($\xi\rightarrow\infty$).
Furthermore,
$$\eqalign{
g_{\nu}^{\alpha}&=\Delta_{\nu\mu}^{i}(q){[\Delta_{i}^{-1}(q)]}^{\mu\alpha}\cr
&= i\{\Delta_{\nu\mu}^{i}(q)[(q^{2}-M^{2}_{i})g^{\mu\alpha}
- q^{\mu}q^{\alpha}] - q_{\nu}q^{\alpha}D^{i}(q)\}~~,\cr}
\EQN Id2$$
and
$$
iq_{\mu}= q^{2}D_{i}(q)q_{\mu}+ M^{2}_{i}q^{\nu}\Delta_{\nu\mu}^{i}(q)~~.
\EQN Id3$$

\vskip 0.5cm

The application of the PT in a theory with non-conserved currents,
such as the SM, is significantly more involved than in the QCD case
\cite{Klako} \cite{D&S}.
The main reasons are the following:

{}~~(a)~~ The charged $W$
couples to fermions with different masses, and
consequently the
elementary Ward identity of \Eq{BasicPinch} gets modified to:
$$\eqalign{
\slashchar{k} &= (\slashchar{p}+
\slashchar{k}-m_{1})-(\slashchar{p}-m_{2}) + (m_{1}-m_{2})\cr
=& S^{-1}(p+k)-S^{-1}(p) + (m_{1}-m_{2}) ~~.\cr}
\EQN BrokenPinch$$
The first two terms of \Eq{BrokenPinch} will pinch and vanish on shell,
respectively, as they did before. But in addition, a term proportional to
$m_{1}-m_{2}$ is left over.
In a general $R_{\xi}$ gauge such terms give rise to extra propagator
and vertex-like contributions, not present in the symmetric case
($m_{1}=m_{2}$).

{}~~(b)~~ Additional graphs involving the "unphysical" Goldstone bosons
$\phi^{+},\phi^{-}$ (which do not couple to the external
fermions if their masses are equal), and physical Higgs
(which does not couple to massless fermions),
must now be included.
Such graphs give rise to new pinch contributions,
even in the Feynman gauge, due to the momenta carried by
interaction vertices such as $\gamma\phi^{+}\phi^{-}$,
$Z\phi^{+}\phi^{-}$, $W^{+}\phi^{-}\phi_{z}$, and $HW^{+}\phi^{-}$.
So, for example, the graphs of Fig.7(d,e,f,g,h) give rise to new
propagator-like pinch contributions.

{}~~(c)~~After the pinch contributions have been identified, particular
care is needed in deciding how to allot them among the
(eventually $\xi$ independent)
quantities one is attempting to construct.
So, unlike the symmetric case ($m_{1}=m_{2}$),
when all propagator-like pinch contributions
were added to the only available self-energy
${\Pi}_{\mu\nu}$, in order to construct the g.i.
${\hat{\Pi}}_{\mu\nu}$, in the case at hand
such pinch contributions must in general
be split among ${\Pi}_{\mu\nu}$, $\Theta_{\mu}$,$\Theta_{\nu}$
and $\Omega$. This is equivalent to saying that, when forming the
inverse of the $W$ self-energy, the longitudinal parts may no longer be
discarded from the four-fermion amplitude, since the charged current
is not conserved, up to terms proportional to the mass splitting.

To concretely address the above three points,
we concentrate on the
propagator-like pinch contributions originating from
vertex and box graphs.
Exactly analogous arguments apply for vertex-like pinch contributions
originating from box graphs.
We define $\Gamma_{0}^{\mu}=
\frac{g}{2\sqrt{2}}\gamma^{\mu}(1-\gamma_{5})$ and
$\Lambda_{0}=
\frac{g}{2M_{w}\sqrt{2}}[m_{1}(1-\gamma_{5})-m_{2}(1+\gamma_{5})$;
when sandwiched between on shell external spinors
$u_{1}(p_{1})$ and $u_{2}(p_{2})$, with $q=p_{1}-p_{2}$ the identity
${\bar{u}}_{1}q_{\mu}\Gamma_{0}^{\mu} u_{2}=
{\bar{u}}_{1}M_{w}\Lambda_{0} u_{2}$ holds.
 There are two types of vertex graphs, those
with an incoming W [graphs of type (a), shown in Fig.7(a, b, c)~],
and those with an incoming $\phi$
[graphs of type (b), shown in Fig.7(d, e, f, g, h)~].
We will call $Q$ and $R$ respectively their
collective propagator-like contributions to the S-matrix.
It is easy to verify that, for any value of $\xi_{i}$,
the most general pinch contribution of any
individual graph of type (a) can be cast in the form
$[Ag^{\mu}_{\rho}+ B \frac{q^{\mu} q_{\rho}}{q^{2}}]
\Gamma^{\rho}_{0}$, whereas
the most general form of an individual
vertex graphs of type (b) is simply $C\Lambda_{0}$.
The factors $A$, $B$, and $C$
depend in general on $q^{2}$,
$M_{w}$, $M_{z}$, the mass of the Higgs $M_{h}$, and $\xi_{i}$,
but {\sl not} on the external fermion momenta $p_{i}$ or any
of the fermion masses $m_{i}$.
Consequently,
$Q$ is of the form (we do not explicitly indicate the external spinors)
$$\eqalign{
Q&=\Gamma^{\sigma}_{0}\Delta_{\sigma\mu}
[{\cal A}g^{\mu}_{\rho}+ {\cal B} \frac{q^{\mu} q_{\rho}}{q^{2}}]
\Gamma^{\rho}_{0}\cr
&=Q_{{\cal A}}+Q_{{\cal B}}\cr}
\EQN Q$$
with ${\cal A}=\sum_{i} A_{i}$,~~ ${\cal B}=\sum_{i} B_{i}$,
and the sum includes all vertex graphs
of type (a). Of course, some of the
individual $A_{i}$ and $B_{i}$ in the above sums may vanish.
Using the identities given in \Eq{Id2} and \Eq{Id3} we obtain for
$Q_{{\cal A}}$ and $Q_{{\cal B}}$:
$$\eqalign{
Q_{{\cal A}}&=
\Gamma^{\sigma}_{0}
\Delta_{\sigma\mu}{\cal A} g^{\mu}_{\rho}
\Gamma^{\rho}_{0}\cr
&=\Gamma^{\sigma}_{0}
\Delta_{\sigma\mu}{\cal A}
{(\Delta^{-1})}^{\mu\nu}\Delta_{\nu\rho}
\Gamma^{\rho}_{0}\cr
&=i\Gamma^{\sigma}_{0}\Delta_{\sigma\mu}
\{[(q^{2}-M^{2}_{i})g^{\mu\nu}- q^{\mu}q^{\nu}]{\cal A}\}
\Delta_{\nu\rho}\Gamma^{\rho}_{0}
-i\Gamma^{\sigma}_{0}\Delta_{\sigma\mu}\{q^{\mu}M_{w}{\cal A}\}D
\Lambda_{0}\cr}
\EQN QA$$
and
$$\eqalign{
Q_{{\cal B}}&=\Gamma^{\sigma}_{0}
\Delta_{\sigma\mu}{\cal B}
\frac{q^{\mu} q_{\rho}}{q^{2}}\Gamma^{\rho}_{0}\cr
&=\Gamma^{\sigma}_{0}
\Delta_{\sigma\mu}{\cal B}
\frac{q^{\mu}}{iq^{2}}[q^{2}Dq_{\rho}+M^{2}_{w}q^{\nu}\Delta_{\nu\rho}]
\Gamma^{\rho}_{0}\cr
&=-i\Gamma^{\sigma}_{0}\Delta_{\sigma\mu}
[M^{2}_{w}{\cal B}\frac{q^{\mu}q^{\nu}}{q^{2}}]
\Delta_{\nu\rho}\Gamma^{\rho}_{0}-i
\Gamma^{\sigma}_{0}\Delta_{\sigma\mu}\{q^{\mu}M_{w}{\cal B}D
\Lambda_{0}\}\cr}
\EQN QB$$

Similarly, for $R$ we have:
$$\eqalign{
R&=\Lambda_{0} D{\cal C}
\Lambda_{0}\cr
&=\Lambda_{0} D{\cal C}
q_{\rho}[\frac{\Gamma^{\rho}_{0}}{M_{w}}]\cr
&=
-i\Lambda_{0} D\{q^{\nu}M_{w}{\cal C}\}\Delta_{\nu\rho}\Gamma^{\rho}_{0}
-i\Lambda_{0}D\{q^{2}{\cal C}\}D\Lambda_{0}\cr}
\EQN R$$

So, the pinch contribution of graphs of type (a) and (b) to
${\hat{\Pi}}_{\mu\nu}$, ${\hat{\Theta}}_{\mu}$,
and $\hat{\Omega}$ are as follows:
$$\eqalign{
{\Pi}_{\mu\nu}^{P}|_{ver}&=
2[g_{\mu\nu}(q^{2}-M^{2}_{w}){\cal A}-
\frac{q_{\mu}q_{\nu}}{q^{2}}(q^{2}{\cal A}+M^{2}_{w}{\cal B})]\cr
{\Theta}_{\mu}^{P}|_{ver}&= -q_{\mu}M_{w}[{\cal A}+{\cal B}+{\cal C}]\cr
{\Theta}_{\nu}^{P}|_{ver}&=-q_{\nu}M_{w}[{\cal A}+{\cal B}+{\cal C}]\cr
{\Omega}^{P}|_{ver}&=-2iq^{2}{\cal C}\cr}
\EQN Allot$$

Similarly, there are two types of box diagrams contributing
propagator-like pinch parts in a general $\xi$ gauge, namely graphs
having no internal scalar particle (Fig.1e),
and graphs containing one internal scalar particle
(Goldstone boson or physical Higgs, Fig.6e ).
If we call their propagator-like
contributions to the amplitude as
$U$ and $K$ respectively, we have
(again omitting external spinors)
$$\eqalign{
&U=\Gamma^{\sigma}_{0}
[{\cal N}g_{\sigma\rho}]\Gamma^{\rho}_{0}~~,\cr
&K= \Gamma^{\sigma}_{0}[\frac{q_{\sigma} q_{\rho}}{q^{2}} {\cal H}]
\Gamma^{\rho}_{0}=
\Lambda_{0}[\frac{M^{2}_{w}}{q^{2}}{\cal H}]\Lambda_{0}~~,\cr}
\EQN Kati$$
where  ${\cal N}=\sum_{i} N_{i}$ and  ${\cal H}=\sum_{j} H_{j}$.
The individual $N_{i}$ and $H_{j}$, some of which may be zero
in a given gauge, depend
in general on $q^{2}$,
$M_{w}$, $M_{z}$,$M_{h}$, and $\xi_{i}$,
but {\sl not} on the external fermion momenta $p_{i}$ or any
of the fermion masses $m_{i}$.
Using \Eq{Id2} we obtain:
$$\eqalign{
U&=
\Gamma^{\sigma}_{0}
g_{\sigma\lambda}{\cal N} g^{\lambda}_{\rho}
\Gamma^{\rho}_{0}\cr
&=\Gamma^{\sigma}_{0}
\Delta_{\sigma\mu}{(\Delta^{-1})}^{\mu}_{\lambda}[{\cal N}]
{(\Delta^{-1})}^{\lambda\nu}\Delta_{\nu\rho}
\Gamma^{\rho}_{0}\cr
&=-\Gamma^{\sigma}_{0}
\{\Delta_{\sigma\mu}[(q^{2}-M^{2}_{w})g^{\mu\lambda}
- q^{\mu}q^{\lambda}] - q_{\sigma}q^{\lambda}D\}[{\cal N}]\cr
&~~~\times\{[(q^{2}-M^{2}_{i})g^{\lambda\nu}- q^{\lambda}q^{\nu}]
\Delta_{\nu\rho} - q_{\lambda}q^{\rho}D\}\Gamma^{\rho}_{0}\cr
&= -\Gamma^{\sigma}_{0}\Delta_{\sigma\mu}
[{(q^{2}-M^{2}_{w})}^{2}{\cal N}g^{\mu\nu}
- q^{\mu}q^{\nu}(q^{2}-2M^{2}_{w}){\cal N}]
\Delta_{\nu\rho}\Gamma^{\rho}_{0}\cr
&~~~-
\Lambda_{0}D[{M^{3}_{w}\cal N}q^{\nu}]\Delta_{\nu\rho}\Gamma^{\rho}_{0}
-\Gamma^{\sigma}_{0}\Delta_{\sigma\mu}
[{M^{3}_{w}\cal N}q^{\mu}]\Lambda_{0}\cr
&~~~-\Lambda_{0}D[q^{2}M^{2}_{w}{\cal N}]D\Lambda_{0}\cr}
\EQN UL$$
and
$$\eqalign{
K&= \Gamma^{\sigma}_{0}q_{\sigma}[\frac{{\cal H}}{q^{2}}]
q_{\rho}\Gamma^{\rho}_{0}\cr
&=-\Gamma^{\sigma}_{0}
[q^{2}Dq_{\sigma}+M^{2}_{w}q^{\mu}\Delta_{\sigma\mu}]
[\frac{{\cal H}}{q^{2}}][q^{2}Dq_{\rho}+M^{2}_{w}q^{\nu}\Delta_{\nu\rho}]
\Gamma^{\rho}_{0}\cr
&=-\Gamma^{\sigma}_{0}\Delta_{\sigma\mu}[\frac{q^{\mu}q^{\nu}}{q^{2}}
M^{4}_{w}{\cal H}]\Delta_{\nu\rho}\Gamma^{\rho}_{0}
-\Lambda_{0}D[q^{2}M^{2}_{w}{\cal H}]D\Lambda_{0}\cr
&~~- \Lambda_{0}D[q^{\nu}M^{3}_{w}{\cal H}]
\Delta_{\nu\rho}\Gamma^{\rho}_{0}
-\Gamma^{\sigma}_{0}\Delta_{\sigma\mu}
[q^{\mu}M^{3}_{w}{\cal H}]D\Lambda_{0}\cr}
\EQN K$$

So, for the propagator-like contributions from the boxes we have

$$\eqalign{
{\Pi}_{\mu\nu}^{P}|_{box}&=
{-(q^{2}-M^{2}_{w})}^{2}{\cal N}g_{\mu\nu}+
\frac{q_{\mu}q_{\nu}}{q^{2}}
[(q^{2}-2M^{2}_{w})q^{2}{\cal N} - M^{4}_{w}{\cal H}]\cr
{\Theta}_{\mu}^{P}|_{box}&= -q_{\mu}M^{3}_{w}
[{\cal N}+{\cal H}]\cr
{\Theta}_{\nu}^{P}|_{box}&=-q_{\nu}M^{3}_{w}
[{\cal N}+{\cal H}]\cr
{\Omega}^{P}|_{box}&= -q^{2}M^{2}_{w}[{\cal N}+{\cal H}]\cr}
\EQN AllotBox$$

The final expressions for the $\xi$-independent
${\hat{\Pi}}_{\mu\nu}$
${\hat{\Theta}}_{\mu}$,
${\hat{\Theta}}_{\nu}$, and $\hat{\Omega}$
are given by
$$\eqalign{
{\hat{\Pi}}_{\mu\nu}&= {\Pi}_{\mu\nu}+
{\Pi}_{\mu\nu}^{P}|_{ver}
 +{\Pi}_{\mu\nu}^{P}|_{box}\cr
{\hat{\Theta}}_{\mu}&= {\Theta}_{\mu} +
{\Theta}_{\mu}^{P}|_{ver} +{\Theta}_{\mu}^{P}|_{box}\cr
{\hat{\Theta}}_{\nu}&= {\Theta}_{\nu} +
{\Theta}_{\nu}^{P}|_{ver} +{\Theta}_{\nu}^{P}|_{box}\cr
\hat{\Omega}&= {\Omega} +
{\Omega}^{P}|_{ver}+{\Omega}^{P}|_{box}~~,\cr}
\EQN General$$
where ${\Pi}_{\mu\nu}$, ${\Theta}_{\mu}$, and ${\Omega}$ are the usual
$WW$, $W\phi$, and $\phi\phi$~ self-energies in an arbitrary gauge $\xi$,
given by the Feynman graphs of Fig.3 , Fig.4 , and Fig.5 , respectively.

Having presented the most general procedure for extracting
$\xi$-independent
${\hat{\Pi}}_{\mu\nu}$
${\hat{\Theta}}_{\mu}$,
${\hat{\Theta}}_{\nu}$, and $\hat{\Omega}$
starting with
any value for the gauge parameters $\xi_{i}$, we can now choose
a convenient gauge to carry out the calculations. Obviously, the
Feynman gauge (with $\xi_{w}=\xi_{z}=\xi_{\gamma}=1$) is by far the most
convenient. In this gauge there are {\sl no}
pinch contributions originating from
boxes ($N_{i}=0$ and $H_{i}=0$ for every $i$ and $j$).
In addition, $B_{i}=0$, for every $i$ in \Eq{Allot}. So, we have:
$$\eqalign{
{\hat{\Pi}}_{\mu\nu}&= {\Pi}_{\mu\nu}^{(\xi = 1)}+
{\Pi}_{\mu\nu}^{P}|_{ver}^{(\xi = 1)}\cr
{\hat{\Theta}}_{\mu}&= {\Theta}_{\mu}^{(\xi = 1)}+
{\Theta}_{\mu}^{P}|_{ver}^{(\xi = 1)}\cr
{\hat{\Theta}}_{\nu}&= {\Theta}_{\nu}^{(\xi = 1)}+
{\Theta}_{\mu}^{P}|_{ver}^{(\xi = 1)}\cr
\hat{\Omega}&= {\Omega}^{(\xi =1)}+
{\Omega}^{P}|_{ver}^{(\xi = 1)}\cr}
\EQN Simp$$

The previous arguments can easily be generalized to construct g.i.
$We\bar{\nu}$ and $\phi\bar{\nu}$
vertices. In the Feynman gauge all one needs to do is
consider all conventional vertex graphs and subtract
from them all propagator-like pinch
contributions, which will be assigned to the self-energy graphs
as described above.
The remaining purely vertex-like structures furnish the g.i. answer.
It is interesting to notice that in any other gauge $\xi\not=1$ one
also needs to consider vertex-like contributions from box diagrams.
In such a case arguments analogous to
the ones presented in this section
will determine how a given vertex-like contribution will be allotted
among the $We\bar{\nu}$ and $\phi e\bar{\nu}$ vertices.

\section {An explicit calculation.}
We will now apply the procedure described above to a simple example.
In this section we prove \Eq{ General}
for the self-energy parts that have an {\sl explicit} dependence
on the Higgs mass $M_{h}$. After
the Higgs dependent propagator-like
pinch contributions from vertices and boxes have been
identified and incorporated into the self energies,
the resulting Higgs dependent terms should form a
g.i. subset of the entire calculation. Even though, as we mentioned already,
when using the PT
one may carry out calculations in
 any convenient gauge (in particular $\xi=1$), we will retain
$\xi$ as a free parameter, in order to demonstrate the numerous
cancellations which take place.
We use the identity
$$
\frac{1}{k^{2}-\xi M^{2}}= \frac{1}{k^{2}-M^{2}}-
\frac{(1-\xi)M^{2}}{(k^{2}-M^{2})(k^{2}-\xi M^{2})}
\EQN Id3$$
to decompose $\xi$-dependent propagators into their
Feynman gauge expressions [first term in the rhs of \Eq{Id3}]
plus an additional $\xi$-dependent piece [second term in the rhs
of \Eq{Id3}],
and we define
$$
F_{1}= \frac{1}{(k^{2}-M_{w}^{2})[{(k+q)}^{2}-M_{h}^{2}]}
\EQN Defa$$
and
$$
F_{2}= \frac{(1-\xi_{w})}{(k^{2}-M_{w}^{2})(k^{2}-\xi_{w} M_{w}^{2})
[{(k+q)}^{2}-M_{h}^{2}]}
\EQN DefF$$

In what follows we omit a common factor
${(\frac{g}{2})}^{2}\int [\frac{d^{4}k}{(2\pi)^{4}}]$.
The Higgs dependent contributions from the
usual self-energy graphs, listed individually, are as follows:

For $\Pi_{\mu\nu}$:
$$\eqalign{
[3f]&= -4M^{2}_{w}g_{\mu\nu}F_{1}+4M^{2}_{w}k_{\mu}k_{\nu}F_{2}\cr
&=[3f]_{(\xi = 1)}+4M^{2}_{w}k_{\mu}k_{\nu}F_{2}\cr}
\EQN P1$$
$$\eqalign{
[3g]&= {(2k+q)}_{\mu}{(2k+q)}_{\nu}F_{1}
-M^{2}_{w}{(2k+q)}_{\mu}{(2k+q)}_{\nu}F_{2}\cr
&=[3g]_{(\xi = 1)}
-M^{2}_{w}{(2k+q)}_{\mu}{(2k+q)}_{\nu}F_{2}\cr}
\EQN P2$$

For $\Theta_{\mu}$:
$$\eqalign{
[4e]&= 2M_{w}{(2k+q)}_{\mu}F_{1}
+2M_{w}k_{\mu}[(2q+k)k]F_{2}\cr
&=[4e]_{(\xi = 1)}+2M_{w}k_{\mu}[(2q+k)k]F_{2}\cr}
\EQN P3$$
$$\eqalign{
[4f]&= -\frac{M^{2}_{h}}{M_{w}}{(2k+q)}_{\mu}F_{1}
-2M^{2}_{h}M_{w}{(2k+q)}_{\mu}F_{2}\cr
&=[4f]_{(\xi = 1)}-M^{2}_{h}M_{w}
{(2k+q)}_{\mu}F_{2}\cr}
\EQN P4$$
Finally, for $\Omega$ we have
\reference{Tadpoles}
{}From the tadpoles shown in Fig.6, only the
one with the $\phi^{+}\phi^{-}$ loop, e.g. (6j) depends
both on $M_{h}$ and $\xi$. In the tadpole (6k)
the contribution
of the Higgs propagator is $\sim\frac{1}{-M_{h}^{2}}$ and cancels
against the contribution from the $H\phi_{z}\phi_{z}$ vertex
which is $\sim M_{h}^{2}$. The remaining expression
depends on $\xi$ and $M_{z}$, and will be combined with other
such terms, not considered in this section.
Finally, the tadpole graph with a Higgs loop (6l) has no
$\xi$-dependence
\endreference
:
$$\eqalign{
[5e]&= -[(2q+k)(2q+k)]F_{1}
+[(2q+k)k][(2q+k)k]F_{2}\cr
&=[5e]_{(\xi = 1)} +[(2q+k)k][(2q+k)k]F_{2}\cr}
\EQN P5$$
$$\eqalign{
[5f]&=\frac{M^{4}_{h}}{M^{2}_{w}}F_{1} -M^{4}_{h}F_{2}\cr
&=[5f]_{(\xi = 1)}-M^{4}_{h}F_{2}\cr}
\EQN P6$$
$$\eqalign{
[5j]&=-\frac{M^{2}_{h}}{M^{2}_{w}}\frac{1}{k^{2}-M^{2}_{w}}
-\frac{M^{2}_{w}}{(k^{2}-M^{2}_{w})(k^{2}-\xi M^{2}_{w})}\cr
&=[5j]_{(\xi = 1)}-\frac{M^{2}_{w}}
{(k^{2}-M^{2}_{w})(k^{2}-\xi M^{2}_{w})}\cr}
\EQN P7$$
Similarly, from the vertex and box graphs of Fig.6 we obtain after
pinching:
$$
[6b]=2M_{w}k_{\mu}F_{2}\Lambda_{0}~~,
\EQN P8$$
$$\eqalign{
[6d]&=F_{1}\Lambda_{0}-[(2q+k)k]F_{2}\Lambda_{0}\cr
&=[6d]_{(\xi = 1)}-[(2q+k)k]F_{2}\Lambda_{0}~~,\cr}
\EQN P9$$
and
$$
[6f]= \Lambda_{0}F_{2}\Lambda_{0}~~.
\EQN P10$$
Notice that the vertex diagram (6c) contains Higgs dependent pinch
contributions even in the Feynman gauge (where the longitudinal
part of the bare $W$ propagator vanishes), due to the momenta carried
by the $WH\phi$ vertex. On the other hand, (6a) and (6e) do not
contribute in the Feynman gauge; indeed $F_{2}(\xi_{w} =1)=0$ in
\Eq{P8} and \Eq{P10}.

In order to cast our results into the form of
\Eq{Q} and \Eq{Kati}
we introduce $F_{3}$, defined through the equation
$$
\int [\frac{d^{4}k}{(2\pi)^{4}}]k_{\mu}F_{2}=
q_{\mu}\int [\frac{d^{4}k}{(2\pi)^{4}}]F_{3} ~~,
\EQN DefPro$$
which projects out $q_{\mu}$,
and arrive to the following expressions:
$$\eqalign{
{\Pi}_{\mu\nu}^{P}|_{ver}^{h}&= 4q_{\mu}q_{\nu}M_{w}^{2}F_{3}\cr
{\Theta}_{\mu}^{P}|_{ver}^{h}&=
q^{\mu}M_{w}\{[6d]_{(\xi = 1)}+
2q^{2}F_{3}+ [(2q+k)k]F_{2}\}\cr
{\Omega}^{P}|_{ver}^{h}&=2q^{2}\{[6d]_{(\xi = 1)}+[(2q+k)k]F_{2}\}\cr}
\EQN AllotHiggs$$
and
$$\eqalign{
{\Pi}_{\mu\nu}^{P}|_{box}^{h}&= q_{\mu}q_{\nu}M_{w}^{2}F_{2}\cr
{\Theta}_{\mu}^{P}|_{box}^{h}&= q_{\mu}q^{2}M_{w}F_{2}\cr
{\Omega}^{P}|_{box}^{h}&= q^{4}F_{2}\cr}
\EQN AllotHiggsBox$$
It is straightforward to verify now that when the contributions
of \Eq{AllotHiggs} and \Eq{AllotHiggsBox} are added to the usual
propagator graphs, e.g. \Eq{P1} through \Eq{P7}, all dependence on
$\xi$ cancels, and
we readily find for the $M_{h}$-dependent parts
of ${\hat{\Pi}}_{\mu\nu}$, ${\hat{\Theta}}_{\mu}$
and $\hat{\Omega}$
the following expressions:
$$\eqalign{
&{\hat{\Pi}}_{\mu\nu}^{h}=[3g]_{(\xi = 1)}+[3h]_{(\xi = 1)}\cr
&{\hat{\Theta}}_{\mu}^{h}=
[4e]_{(\xi = 1)}+[4f]_{(\xi = 1)}+ q^{\mu}M_{w}[6d]_{(\xi = 1)}\cr
&{\hat{\Omega}}^{h}=[5e]_{(\xi = 1)}+[5f]_{(\xi = 1)}
+[5j]_{(\xi = 1)}+ 2q^{2}[6d]_{(\xi = 1)}\cr}
\EQN HiggsFinal$$

\vskip 0.3cm
As anticipated, the final $\xi$-independent results coincide
with the results obtained by calculating
both the self-energies and the pinch graphs at $\xi=1$, which is certainly
the simplest possible gauge choice.
It is interesting to notice that all gauge cancellations
become manifest
through simple algebraic manipulations of the Feynman integrals;
in particular, no integrations over the loop momenta need be
carried out.

\section {Ward identities and S-matrix rearrangement}

In the previous section we showed explicitly how the application of the PT
gives rise to g.i. self energies, mixing terms, and vertices.
We now proceed to investigate the Ward identities these quantities satisfy.
In deriving these
Ward identities, knowledge of the explicit
form of the quantities involved
is not required. All one needs is the fact that these quantities
have been rendered g.i. through the PT,
and the requirement that the residual $\xi$-dependence,
stemming from the longitudinal part of the tree-level $W$ propagator
(\Eq{Wprop}) as well as
the tree-level $\phi$ propagator (\Eq{Gold}), cancels out
from each of the S-matrix
amplitudes ${\hat{T}}_{1}$, ${\hat{T}}_{2}$, and ${\hat{T}}_{3}$.

To see this explicitly we first concentrate on the ${\hat{T}}_{2}$ part of the
S-matrix
\reference{Mirors}
There is of course the mirror contribution to ${\hat{T}}_{2}$, with the
vertex hooked on the upper part of Fig.2(e, f), which can be
manipulated in exactly the same way
\endreference
{}.
 Using \Eq{Id1}, and omitting the external spinors,
we find
$$\eqalign{
{\hat{T}}_{2}&= \Gamma^{\sigma}_{0}
\Delta_{\sigma\mu}{\hat{\Gamma}}^{\mu}
+\Lambda_{0} D_{\xi}\hat{\Lambda}\cr
&=\Gamma^{\sigma}_{0}
[{\cal D}_{\sigma\mu}+ D_{\xi}\frac{q_{\sigma}q_{\mu}}{M^{2}}]
{\hat{\Gamma}}^{\mu}+\Lambda_{0}D_{\xi}\hat{\Lambda}\cr}
\EQN T2$$
where $D_{\xi}$ is given by \Eq{Gold}. We have added the subscript $\xi$ to
explicitly indicate that these terms contain
the entire residual $\xi$ dependence of
the ${\hat{T}}_{2}$ amplitude,
and have removed the
subscript "$w$" from the mass of the $W$ to avoid notational clutter.
 The requirement of $\xi$-independence
of ${\hat{T}}_{2}$ readily yields:
$$
[\frac{\Lambda_{0}}{M}]D_{\xi}
(q^{\mu}{\hat{\Gamma}}_{\mu}-M\hat{\Lambda})=0
\EQN VCond$$
for each value of $\xi$, from which immediately follows that
${\hat{\Gamma}}_{\mu}$ and $\hat{\Lambda}$ must be related by the
following Ward identity:
$$
q^{\mu}{\hat{\Gamma}}_{\mu}=M\hat{\Lambda}~~,
\EQN VWI$$
which is the one-loop generalization of the tree-level Ward identity
$q^{\mu}\Gamma^{\mu}_{0}=M_{w}\Lambda_{0}$.
The $\xi$-independent ${\hat{T}}_{2}$ is then given by
$$
{\hat{T}}_{2}= \Gamma^{\nu}_{0}[g_{\mu\nu}-\frac{q_{\nu}q_{\mu}}{M^{2}}]
\frac{{\hat{\Gamma}}^{\mu}}{q^{2}-M^{2}}
\EQN RRA$$

We now proceed to repeat a similar argument for ${\hat{T}}_{1}$.
We have:
$$\eqalign{
{\hat{T}}_{1}&= \Gamma^{\sigma}_{0}
\Delta_{\sigma\mu}{\hat{\Pi}}^{\mu\nu}
\Delta_{\nu\rho}\Gamma^{\rho}_{0}
+ \Lambda_{0} D\hat{\Omega} D\Lambda_{0}\cr
&+ \Gamma^{\sigma}_{0}
\Delta_{\sigma\mu}{\hat{\Theta}}^{\mu}D\Lambda_{0}
+\Lambda_{0} D{\hat{\Theta}}^{\nu}\Delta_{\nu\rho}\Gamma^{\rho}_{0}\cr}
\EQN T1$$
or after using \Eq{Id1}
$$\eqalign{
{\hat{T}}_{1}&= \Gamma^{\sigma}_{0}
[{\cal D}_{\sigma\mu}+ \frac{q_{\sigma}q_{\mu}}{M^{2}}D_{\xi}]
{\hat{\Pi}}^{\mu\nu}
[{\cal D}_{\nu\rho}+ \frac{q_{\nu}q_{\rho}}{M^{2}}D_{\xi}]
\Gamma^{\rho}_{0}
+ \Lambda_{0} D_{\xi}\hat{\Omega} D_{\xi}\Lambda_{0} \cr
&+ \Gamma^{\sigma}_{0}
[{\cal D}_{\sigma\mu}+ \frac{q_{\sigma}q_{\mu}}{M^{2}}D_{\xi}]
{\hat{\Theta}}^{\mu}D_{\xi}\Lambda_{0}
+\Lambda_{0} D_{\xi}{\hat{\Theta}}^{\nu}
[{\cal D}_{\nu\rho}+ \frac{q_{\nu}q_{\rho}}{M^{2}}D_{\xi}]
\Gamma^{\rho}_{0}\cr}
\EQN PCon$$
Demanding that ${\hat{T}}_{1}$ is $\xi$-independent we obtain:
$$
q^{\nu}q^{\mu}{\hat{\Pi}}_{\mu\nu}
-2Mq^{\mu}{\hat{\Theta}}_{\mu}+M^{2}\hat{\Omega}=0 ~~,
\EQN PWI$$
and
$$
q^{\mu}{\hat{\Pi}}_{\mu\nu}-M{\hat{\Theta}}_{\nu}=0~~.
\EQN PWI1$$
{}From \Eq{PWI} and \Eq{PWI1} follows that:
$$
q^{\nu}q^{\mu}{\hat{\Pi}}_{\mu\nu}- M^{2}\hat{\Omega}=0 ~~,
\EQN PWI2$$
and
$$
q^{\mu}{\hat{\Theta}}_{\mu}- M\hat{\Omega}=0 ~~.
\EQN PWI3$$
Finally, the g.i. ${\hat{T}}_{1}$ is given by
$$
{\hat{T}}_{1}=\Gamma^{\sigma}_{0}{\cal D}_{\sigma\mu}{\hat{\Pi}}^{\mu\nu}
{\cal D}_{\nu\rho}\Gamma^{\rho}_{0} ~~.
\EQN RRB$$

The next step in our analysis is to use the Ward identities derived above,
to reformulate the S-matrix in terms of individually g.i.
transverse and longitudinal pieces. Such a reformulation
may be thought of as giving rise to a
new transverse g.i. $W$ self-energy
${\hat{\Pi}}_{\mu\nu}^{t}$
and a new transverse vertex ${\hat{\Gamma}}_{\mu}^{t}$,
 with a
 g.i.longitudinal part left over. The cost of such a reformulation
is the appearance of massless Goldstone poles in our expressions. However,
since both the old and the new quantities originate from the same
{\sl unique} S-matrix, all poles introduced by this reformulation cancel
against each other, because the S-matrix contains no massless poles to
begin with.

We start with ${\hat{T}}_{2}$.
We define
$$\eqalign{
{\hat{\Gamma}}_{\mu}^{t}&=[g_{\mu\nu}-\frac{q_{\mu}q_{\nu}}{q_{2}}]
{\hat{\Gamma}}^{\nu}\cr
&= {\hat{\Gamma}}_{\mu}+ \frac{q_{\mu}}{q^{2}}M\hat{\Lambda}\cr}
\EQN VertDef$$
with
$$
q^{\mu}{\hat{\Gamma}}_{\mu}^{t}=S^{-1}_{i}(p_{1})-S^{-1}_{j}(p_{2})=0
\EQN OnShWI$$
for on-shell external fermions.
Using the identity
$$
\frac{1}{M^{2}}= \frac{1}{q^{2}}+ \frac{q^{2}-M^{2}}{q^{2}M^{2}}
\EQN AnotherId$$
we obtain
$$\eqalign{
{\hat{T}}_{2}&= \Gamma^{\sigma}_{0}
[g_{\sigma\mu}-\frac{q_{\sigma}q_{\mu}}{q^{2}}-
\frac{(q^{2}-M^{2})q_{\sigma}q_{\mu}}{q^{2}M^{2}}]
\frac{1}{q^{2}-M^{2}}{\hat{\Gamma}}^{\mu} \cr
&=\Gamma^{\sigma}_{0}
[g_{\sigma\mu}-\frac{q_{\sigma}q_{\mu}}{q^{2}}]
\frac{1}{q^{2}-M^{2}}
{\hat{\Gamma}}^{\mu}
-\Lambda_{0} \frac{1}{q^{2}}
\hat{\Lambda}\cr
&=\Gamma^{\sigma}_{0}
[\frac{g_{\sigma}^{\mu}}{q^{2}-M^{2}}]
{\hat{\Gamma}}_{\mu}^{t}
-\Lambda_{0}\frac{1}{q^{2}}\hat{\Lambda}\cr}
\EQN Ref1$$
It is now obvious that ${\hat{T}}_{2}$ in \Eq{Ref1} can be thought of as
consisting of two different pieces, one due to a massive vector boson
of mass $M$, and one due to a massless Goldstone boson.

Similarly, if we write ${\hat{\Theta}}_{\mu}$ in the form
$$
{\hat{\Theta}}_{\mu}=q_{\mu}\hat{\Theta}
\EQN TensProp$$
with
$$
\hat{\Theta}= \frac{M\hat{\Omega}}{q^{2}}
\EQN TTT$$
from \Eq{PWI3}, we can define the
g.i. quantity ${\hat{\Pi}}_{\mu\nu}^{t}$ in terms of
${\hat{\Pi}}^{\mu\nu}$ and $\hat{\Theta}$ as follows:
$$\eqalign{
{\hat{\Pi}}_{\mu\nu}^{t}&=
{\hat{\Pi}}_{\mu\nu}-
\frac{q_{\mu}q_{\nu}}{q^{2}}M\hat{\Theta}\cr
&=(g_{\mu\sigma}-\frac{q_{\mu}q_{\sigma}}{q^{2}}){\hat{\Pi}}^{\sigma\rho}
(g_{\nu\rho}-\frac{q_{\nu}q_{\rho}}{q^{2}})\cr}
\EQN TransProp$$
where \Eq{PWI1} and \Eq{TTT} have been used.
${\hat{\Pi}}_{\mu\nu}^{t}$ is transverse, e.g.
$$
q^{\mu}{\hat{\Pi}}_{\mu\nu}^{t}=q^{\nu}{\hat{\Pi}}_{\mu\nu}^{t}=0
\EQN TransWI$$
We may now re-express ${\hat{T}}_{1}$ of \Eq{RRB} in terms of
${\hat{\Pi}}_{\mu\nu}^{t}$ and $\hat{\Omega}$
as follows:
$$\eqalign{
{\hat{T}}_{1}&= {\Gamma}^{\mu}_{0}[\frac{1}{q^{2}-M^{2}}]
({\hat{\Pi}}^{\mu\nu}+
\frac{q_{\mu}q_{\nu}}{q^{2}}M\hat{\Theta})[\frac{1}{q^{2}-M^{2}}]
{\Gamma}^{\nu}_{0}
+\Lambda_{0}\frac{1}{q^{2}}\hat{\Omega}\frac{1}{q^{2}}\Lambda_{0}\cr
&= {\Gamma}^{\mu}_{0}[\frac{1}{q^{2}-M^{2}}]{\hat{\Pi}}_{\mu\nu}^{t}
[\frac{1}{q^{2}-M^{2}}]{\Gamma}^{\nu}_{0}+
\Lambda_{0}\frac{1}{q^{2}}\hat{\Omega}\frac{1}{q^{2}}\Lambda_{0}\cr}
\EQN T1Ref$$
Again, ${\hat{T}}_{1}$ of \Eq{T1Ref} is the sum of two self-energies, one
corresponding to a regular massive vector field and one to a massless
Goldstone boson. It is interesting to notice that the above rearrangements
have removed the one-loop mixing terms ${\hat{\Theta}}_{\mu}$ and
${\hat{\Theta}}_{\nu}$ between $W$ and $\phi$, thus leading to a
generalization of the well known tree-level property of the $R_{\xi}$ gauges.
It is important to emphasize again that the massless poles in the above
expressions would not have appeared had we not insisted on the
transversality of the $W$ self-energy and vertex. They are therefore
not related to any particular gauge choice, such as the Landau gauge
($\xi=0$).
The "pure" box contributions which are left over constitute
the g.i. subset of the amplitude called
${\hat{T}}_{3}(s,t,m_{1},m_{2})$ in \Eq{S2-matrix}; it can also be
decomposed into transverse and longitudinal parts, but such a task is
beyond the scope of this paper.

The above reformulation has both conceptual and computational merits.
{}From the point of view of the S.D. equations, it makes
the renormalizability
of the resulting equations manifest.
If we were to extract a S.D. equation for the g.i. $WW$ self-energy
${\hat{\Pi}}^{\mu\nu}$ from \Eq{T1} instead of
\Eq{T1Ref}, our expressions would
be multiplied by the tensor~~~
 $(g_{\mu\nu}-\frac{q_{\mu}q_{\nu}}{M^{2}})$,
and would therefore give
rise to non-renormalizable terms, exactly as it happens in the unitary
gauges. In addition, as explained in the Introduction,
this rearrangement leads to
considerable calculational simplifications, since it organizes the
transverse and longitudinal pieces in individually g.i. blocks.
This is particularly economical if one is only interested in
g.i longitudinal
contributions, as is the case in the context of CP violation in
semileptonic top quark decays \cite{Sonb}.

\section{Proof of the Ward identities at one loop}
We now proceed to explicitly prove the validity of
\Eq{PWI} and \Eq{PWI3} to one-loop.
As
Passarino and Veltman showed
\cite{P&V},
 the conventional one-loop gauge
self-energies, calculated in the
Feynman gauge, satisfy the following Ward identity
$$
q^{\mu}q^{\nu}{\Pi}^{(\xi = 1)}_{\mu\nu} -
2M_{w}q^{\mu}\Theta_{\mu}^{(\xi = 1)}+ M^{2}_{w}\Omega^{(\xi = 1)}
= 0 ~~.
\EQN P&V$$
In fact, \Eq{P&V} holds for all values of $\xi$, as one can derive
from the Slavnov-Taylor identity
$$\eqalign{
&<T{\partial}_{\mu}W^{\mu}(x){\partial}_{\nu}W^{\nu}(y)>+
\xi M_{w}<T{\partial}_{\mu}W^{\mu}(x)\phi (y)>\cr
&+\xi M_{w}<T\phi (x){\partial}_{\mu}W^{\mu}(y)>+
{\xi}^{2}M_{w}^{2}<T\phi (x)\phi (y)>=-i\xi\delta(x-y)~~,\cr}
\EQN Jeger$$
relating the corresponding time ordered two-point functions,
which is valid for all values of $\xi$
\reference{Jeger}
F.~Jegerlehner, Lecture notes for TASI 91
\endreference
{}.
It is straightforward to verify
that \Eq{P&V} holds
for the corresponding pinch contributions as well, e.g.
$$
q^{\mu}q^{\nu}{\Pi}_{\mu\nu}^{P}|_{ver}^{(\xi = 1)} -
2M_{w}q^{\mu}{\Theta}_{\mu}^{P}|_{ver}^{(\xi = 1)}
+ M^{2}_{w}{\Omega}^{P}|_{ver}^{(\xi = 1)} = 0 ~~,
\EQN P&VPinch$$
already at the level of
\Eq{Allot} and \Eq{AllotBox}; indeed, they individually satisfy
\Eq{P&VPinch}, regardless of the explicit closed form of the quantities
${\cal A}$, ${\cal B}$, ${\cal C}$, ${\cal N}$, and ${\cal H}$, for every
value of $\xi$.
Adding \Eq{P&V} and \Eq{P&VPinch} by parts, immediately yields
\Eq{PWI}.

For completeness,
we will now explicitly prove \Eq{P&VPinch}. We work in the
$\xi=1$ gauge.
If we define
$$
I_{ij}(q)= {\mu}^{4-n}\int [\frac{d^{n}k}{(2\pi)^{n}}]
\frac{1}{(k^{2}-M_{i}^{2})[{(k+q)}^{2}-M_{j}^{2}]}~~,
\EQN DefI$$
we have after the pinching the vertex graphs of Fig.7 (with
$\xi_{w}=\xi_{z}=\xi_{\gamma}=1$):
$$
[a+b+c]= {\Gamma}^{\sigma}_{0}\Delta_{\sigma\mu}
[2s^{2}I_{w\gamma}+2c^{2}I_{wz}]{\Gamma}^{\mu}_{0}
\EQN One$$
$$
[d+e+f+g+h]=\Lambda_{0}D\{s^{2}I_{w\gamma}+
[\frac{{(c^{2}-s^{2})}^{2}}{4c^{2}}+ \frac{1}{4}]
I_{wz}+\frac{I_{wh}}{4}\}\Lambda_{0}
\EQN Two$$
Using \Eq{Allot} and introducing
$$
\kappa=\frac{{(c^{2}-s^{2})}^{2}}{4c^{2}}+ \frac{1}{4}~~,
\EQN LF$$
we obtain from \Eq{One} and \Eq{Two}
the following expressions:
$$\eqalign{
&{\Pi}_{\mu\nu}^{P}|_{ver}^{(\xi = 1)}=
4[(q^{2}-M^{2}_{w})g^{\mu\nu}- q^{\mu}q^{\nu}](s^{2}I_{w\gamma}+c^{2}I_{wz})
\cr
&
{\Theta}_{\mu}^{P}|_{ver}^{(\xi = 1)}=
q_{\mu}M_{w}[3s^{2}I_{w\gamma}+
(2c^{2}+\kappa)I_{wz}+\frac{1}{4}I_{wh}]\cr
&
{\Omega}^{P}|_{ver}^{(\xi = 1)}=
2q^{2}[s^{2}I_{w\gamma}+
\kappa I_{wz}+\frac{I_{wh}}{4}]\cr}
\EQN A3$$
and so
$$\eqalign{
&q^{\mu}q^{\nu}{\Pi}_{\mu\nu}^{P}|_{ver}^{(\xi = 1)}=
-4M^{2}_{w}q^{2}(s^{2}I_{w\gamma}+c^{2}I_{wz})\cr
&2M_{w}q^{\mu}{\Theta}_{\mu}^{P}|_{ver}^{(\xi = 1)}=
-2M^{2}_{w}q^{2}[3s^{2}I_{w\gamma}+(2c^{2}+\kappa)I_{wz}
+\frac{1}{4}I_{wh}]\cr
&M^{2}_{w}{\Omega}^{P}|_{ver}^{(\xi = 1)}=
-2M^{2}_{w}q^{2}
[s^{2}I_{w\gamma}+
\kappa I_{wz}
+\frac{I_{wh}}{4}]\cr}
\EQN A4$$
Adding by parts we arrive at the advertised result of \Eq{P&VPinch}

Turning to \Eq{PWI3}, it is interesting to notice
(and elementary to prove) that
\Eq{PWI3} is automatically satisfied by the fermionic contributions
to the conventional $\Theta_{\mu}$ and $\Omega$ self-energies
(tadpoles must be included).
This comes as no surprise, since the fermionic loops are
g.i by themselves, and receive therefore no pinch contributions.
{}From this point of view the effect of
pinching is to "abelianize" the bosonic sector, which, after being rendered
g.i., ends up satisfying
the QED-like Ward identity of \Eq{PWI3} as well.

We now prove \Eq{PWI3} for the bosonic contributions.
To that end we must evaluate $q^{\mu}{\Theta}_{\mu}^{(\xi = 1)}$
and $M_{w}{\Omega}^{(\xi = 1)}$. In doing so we find it more
economical to act directly with $q^{\mu}$ on the
individual graphs that define
${\Theta}_{\mu}^{(\xi = 1)}$, instead of first computing them and
contracting with $q^{\mu}$ the final answer.
{}From the graphs of Fig.4 and Fig.5 we obtain (see also \cite{P&V}):
$$\eqalign{
\frac{q^{\mu}{\Theta}_{\mu}^{(\xi = 1)}}{M_{w}}&=
-(q^{2}-2M_{w}^{2})s^{2}I_{w\gamma}
+[\frac{3}{4}q^{2}-\frac{{(M_{w}^{2}-M_{h}^{2})}^{2}}{M_{w}^{2}}]I_{wh}\cr
&+\Biggl\lbrack
(2c^{2}-3\kappa)q^{2}
+\frac{s^{2}(1+8c^{2})}{4c^{2}}[M_{w}^{2}-M_{z}^{2}]
\Biggr\rbrack
I_{wz}\cr}
\EQN A5$$
and
$$\eqalign{
{\Omega}^{(\xi = 1)}&=
-2(q^{2}-M_{w}^{2})s^{2}I_{w\gamma}
+[q^{2}-\frac{{(M_{w}^{2}-M_{h}^{2})}^{2}}{M_{w}^{2}}]I_{wh}\cr
&-\Biggl\lbrack
2\kappa q^{2}
-\frac{s^{2}(1+8c^{2})}{4c^{2}}[M_{w}^{2}-M_{z}^{2}]
\Biggr\rbrack I_{wz}\cr}
\EQN A6$$

The terms proportional to $I_{w\gamma}$ in the r.h.s. of \Eq{A5}
are the sum of the graphs 4a, 4c, and 4g,
those proportional to $I_{wh}$ are the sum
of 4e and 4f, and those proportional to $I_{wz}$ are the sum
of 4b, 4d, 4h, and 4i; in calculating the latter, we used the
algebraic identity
$\frac{3s^{2}(c^{2}-s^{2})}{4c^{2}}+\frac{c^{2}}{2}=2c^{2}-3\kappa$
. Similarly,
the terms proportional to $I_{w\gamma}$ in the r.h.s. of \Eq{A6}
are the sum of 5a and 5c, those proportional to $I_{wh}$ are the sum
of 5e and 5f, and those proportional to $I_{wz}$ are the sum
of 5b, 5d, 5g, and 5h.
It is now straightforward to verify that \Eq{A5} and \Eq{A6} combined with
the  expressions for $q^{\mu}{\Theta}_{\mu}^{P}|_{ver}^{(\xi = 1)}$ and
${\Omega}^{P}|_{ver}^{(\xi = 1)}$ from \Eq{A4} yield the desired result of
\Eq{PWI3}.
In the previous calculations we omitted seagull-like terms, e.g.
terms proportional to $\int \frac{d^{4}k}{k^{2}-M^{2}_{i}}$, which emerge
every time the elementary identity
$2qk= \{[(k+q)^{2}-M^{2}_{j}]-(k^{2}-M^{2}_{i}) +(M^{2}_{j}-M^{2}_{i})\}$
is used. It is easy to show however that all such terms,
when combined with the regular tadpole and seagull terms
of $\Theta_{\mu}^{(\xi=1)}$ and $\Omega^{(\xi=1)}$, exactly satisfy
\Eq{PWI3}.
It is important to emphasize that the validity of \Eq{PWI3} is particular
to the g.i. self-energies constructed via the PT. Indeed,
as we see directly form \Eq{A4}, \Eq{A5} and \Eq{A6}, neither
${\Theta}_{\mu}^{(\xi = 1)}$ and ${\Omega}^{(\xi = 1)}$, nor
${\Theta}_{\mu}^{P}|_{ver}^{(\xi = 1)}$ and
${\Omega}^{P}|_{ver}^{(\xi = 1)}$ satisfy \Eq{PWI3}. It is only
after they are combined to
form ${\hat{\Theta}}_{\mu}$ and $\hat{\Omega}$ that \Eq{PWI3} holds.
After \Eq{P&VPinch} and \Eq{PWI3}
have been proved, \Eq{PWI1} and \Eq{PWI2} follow immediately.
Following similar steps, it is easy to verify that
${\hat{\Gamma}}_{\mu}$ and $\hat{\Lambda}$
satisfy \Eq{VWI}.

\section{Conclusions}
In this paper we showed how to apply the PT in the case of a
four-fermion amplitude,
with non-conserved external currents.
In particular, we concentrated on the case of the charged currents,
where the non-conservation of the current is due to the
mass difference of the two fermions merging in a $W$-vertex.
We showed how to construct via a
well-defined procedure one-loop g.i. $W^{+}W^{-}$, $W^{+}\phi^{-}$,
$\phi^{+} W^{-}$, and $\phi^{+}\phi^{-}$
self-energies, and g.i. $W\bar{f_{1}}f_{2}$ and
$\phi\bar{f_{1}}f_{2}$ vertices.
In addition we derived a set of QED-like
Ward identities, which these g.i. quantities satisfy.
The Ward identities allow for the
decomposition of the S-matrix
in manifestly g.i. transverse and
longitudinal pieces, with distinct kinematic properties, e.g.
propagator-like, vertex-like, and box-like.

Although throughout this paper we used electrons and neutrinos as
external particles, with $m_{\nu}=0$ and $m_{e}\not= 0$,
our results can be immediately generalized to
the case of
arbitrary fermions with $m_{1}\not= m_{2}$, and $m_{1}m_{2}\not= 0$.
The only additional technical point to be made is that in such a case
(unlike the neutrino) both fermions couple to the photon and
the physical Higgs particle.

 The generalization of the analysis of the present paper to the
case of neutral non-conserved currents is a more laborious task.
{}From the technical point of view the situation is slightly more involved,
because in addition to the $ZZ$, $Z\phi_{z}$, $\phi_{z} Z$ and
$\phi_{z}\phi_{z}$ self-energies and the $Zf\bar{f}$, $\phi_{z}f\bar{f}$
vertices
(which are the equivalent of the
$W^{+}W^{-}$, $W^{+}\phi^{-}$, $\phi^{+} W^{-}$, and $\phi^{+}\phi^{-}$
self-energies, and
 $W\bar{f_{1}}f_{2}$
and $\phi\bar{f_{1}}f_{2}$ vertices we considered here),
we also have the additional $\gamma\gamma$, $\gamma Z$, $Z\gamma$
$\gamma\phi_{z}$, $\phi_{z}\gamma$, $HH$, $HZ$, $ZH$, $\gamma H$,
$H\gamma$, $H\phi_{z}$, and $\phi_{z} H$ self-energies,
and the additional
$\gamma f\bar{f}$ and $Hf\bar{f}$ vertices
to take into account.

There is another important area where the application of the PT
has proven
useful, namely the study of three-boson couplings for the
upcoming LEP~2 process
$e^{+}e^{-}\rightarrow W^{+}W^{-}$.
In \cite{Kostas} g.i. $\gamma W^{+}W^{-}$ $ZW^{+}W^{-}$ vertices
for off-shell $\gamma$ and $Z$
have been constructed, under the assumption that
the external electrons are
massless. The g.i $\gamma W^{+}W^{-}$ vertex
${\hat{\Gamma}}_{\mu\alpha\beta}(q,p,-p-q)$
was shown to satisfy a QED-like Ward identity, relating it to the
g.i. $WW$ self-energies, e.g.
$q^{\mu}{\hat{\Gamma}}_{\mu\alpha\beta}(q,p,-p-q)=
{\hat{\Pi}}_{\alpha\beta}(p)
-{\hat{\Pi}}_{\alpha\beta}(p+q)$.
A similar, more complicated Ward identity holds true for the $ZW^{+}W^{-}$
vertex. In order to derive it, following the method
described in section 4, one needs to
treat the external fermions as
massive, so that the one-loop vertex $\phi_{z} W^{+}W^{-}$,
which is a crucial ingredient of the Ward identity in question,
can appear in the amplitude.
We will report the results of this analysis
elsewhere.

\section {Acknowledgements}
The author thanks K.~Philippides, M.~Schaden, and A.~Sirlin
for useful discussions.
This work was supported by the National Science Foundation under
Grant No.PHY-9017585.

\section {References.}
\ListReferences
\section{Figure Captions}

1.~~Graphs (a)-(c) are some of the contributions to the S-matrix
for four-fermion processes.
Graphs (e) and (f) are pinch parts, which, when added to the usual
self-energy graphs (d), give rise to the gauge-independent amplitude
${\hat{T}}_{1}(t)$.
The mirror image graph corresponding to (b) is not shown.

2.~~ The
g.i. self-energies ${\hat{\Pi}}_{\mu\nu}$,
$\hat{\Omega}$,
${\hat{\Theta}}_{\mu}$
and ${\hat{\Theta}}_{\nu}$ (a, b, c, and d, respectively)
and the g.i. vertices ${\hat{\Gamma}}_{\mu}$ and $\hat{\Lambda}$
(e and f, respectively).

3.~~The Feynman diagrams contributing to the $W$ self-energy. Seagull and
tadpole graphs are not shown. The dotted lines represent the
Faddeev-Popov ghosts.

4.~~The Feynman diagrams for the one-loop $\phi W$ mixing.

5.~~The Feynman diagrams for the $\phi\phi$ self-energy. Notice the absence
of any ghosts corresponding to the photon.

6.~~Vertex and box diagrams containing a Higgs propagator and their
propagator-like pinch contributions.

7.~~ The vertex graphs which give propagator-like pinch contributions
in the Feynman gauge.

\bye